\documentstyle[12pt]{article}

\begin{document}
\begin{center}
SELF-CONSISTENT SEPARABLE RPA FOR DENSITY- \\
AND CURRENT-DEPENDENT FORCES

\vspace{0.5cm}
J. Kvasil$^1$, V.O. Nesterenko$^2$, and P.-G. Reinhard$^3$

\vspace{0.3cm}
$^1$ 
Institute of Particle and Nuclear Physics, Charles
University, V.Hole\v sovi\v ck\'ach 2, CZ-18000 Praha 8, Czech Republic\\
E-mail: kvasil@ipnp.troja.mff.cuni.cz

\vspace{0.3cm}
$^2$
BLTP, Joint Institute for Nuclear Research
141980, Dubna, Moscow Region, Russia \\ E-mail: nester@thsun1.jinr.ru

\vspace{0.3cm}
$^3$
Institut fur Theoretische Physik,Universitat Erlangen,
W-8520 Erlangen, Germany\\
E-mail: mpt218@theorie2.physik.uni-erlangen.de

\vspace{1cm}
ABSTRACT

\end{center}
\vspace{0.3cm}

Self-consistent factorization of two-body
residual interaction is proposed for arbitrary density- and
current-dependent energy functionals. Following this procedure, a
separable RPA (SRPA) method is constructed. SRPA dramatically
simplifies the calculations and demonstrates quick convergence to
exact results.  The method is tested for SkM* forces.

\newpage
\section{Introduction}

Effective nucleon-nucleon interactions (Skyrme, Gogny, ...) are
widely used for description of nuclear properties (see, for
example, Ref.\cite{Rew}). However, their application to nuclear
dynamics is rather limited even in the linear regime. The latter
is usually treated within random-phase-approximation (RPA) and
assumes diagonalization of matrices with the rank determined by
the size of the particle-hole $(ph)$ configuration space. In
deformed and heavy spherical nuclei this space is impressive and
condemns to a huge computational effort. As a result, the
effective forces are mainly applied to study ground state
properties while RPA calculations are very limited.

RPA problem becomes much simpler if the residual two-body
interaction is factorized (reduced to a separable form):
\begin{eqnarray}
\sum_{mnij}&& \!\!\! <mn|V_{res}|ij> a^+_m a^+_n a_j a_i \rightarrow
\sum_{k,k'=1}^{K} \kappa_{k,k'} {\hat X}_k {\hat X}_{k'}, \nonumber \\
&& {\hat X}_k = \sum_{ph} <p|{\hat X}_k |h> a^+_p a_h\; .
\label{1}
\end{eqnarray}
This allows to avoid dealing with high-rank matrices. The
main trouble is to accomplish the factorization self-consistently,
with minimal number of separable terms and high accuracy.

Factorization of the residual interaction is widely used in
nuclear theory but mainly within trivial schemes exploiting one
separable term with an intuitive guess for the separable one-body
operator ${\hat X}$. The strength constant $\kappa$ is usually
fitted so as to reproduce available experimental data (see e.g.
\cite{Sol}). The separable interaction thus obtained does not
depend on nuclear density and is not consistent with the mean
field. Obviously, accuracy and predictability of such schemes are
rather low. Several self-consistent schemes$^{3-8}$
proposed during last decades
signified a certain progress in this problem. However, these
schemes were not sufficiently general. Some of them were limited
to analytic or simple numerical estimates$^{3-6}$
the others were not fully self-consistent \cite{Ne_PRC} or
covered only particular effective forces \cite{Vor}.

In the present paper we propose a general self-consistent
separable RPA (SRPA) approach
relevant to arbitrary density- and current-dependent functionals.
The
self-consistent scheme of Ref.\cite{LS} is
generalized to the case of several separable operators. The
operators have maxima at different slices of the nucleus, which
is crucial for accurate reproduction of $V_{res}$ and high
numerical accuracy.
SRPA is tested for the case of SkM* functional \cite{Rei}.

\section{General formulation of SRPA}

The nucleus is assumed to undergo small-amplitude harmonic
vibrations around HF or HFB ground state. The starting point is a
general time-dependent energy functional $E(J_{s}^{\alpha}({\vec
r},t))$ depending on a set of arbitrary neutron and proton
densities and currents $J_{s}^{\alpha}({\vec r},t)$ ($s=n,p$;
$\alpha$ labels densities and currents)
\begin{equation}
E(J_{s}^{\alpha}({\vec r},t))=
<\Psi(t) |\hat H| \Psi(t)>=\int {\cal H} ({\vec r})d{\vec r}
\label{2}
\end{equation}
where $|\Psi(t)\!>$ is the wave function of the vibrating system
described as the time-dependent Slater determinant.
Time-dependent densities and currents are determined through
the corresponding operators as
\begin{equation}
J_{s}^{\alpha}({\vec r},t)=
<\Psi(t) |\hat J_{s}^{\alpha}({\vec r})| \Psi(t)>.
\label{3}
\end{equation}
The wave function $|\Psi(t)\!>$ is obtained
from the static HF Slater determinant $|\Psi_0\!>$ by the scaling
transformation \cite{LS}
\begin{equation}
|\Psi(t)\!>=\prod_{k=1}^K
exp[-i(q_{sk}(t)-\!<\!\!q_{sk}\!\!>)\hat{P}_{s k}]
exp[-ip_{sk}(t)\hat{Q}_{s k}]
|\Psi_0>
\label{4}
\end{equation}
where ${\hat Q}_{sk}$ and ${\hat P}_{sk}$ are, respectively, T-even
and T-odd generators fulfilling the equations
\begin{eqnarray}
\hat{Q}_{sk}^+=\hat{Q}_{sk},\quad
T\hat{Q}_{sk}T^{-1}=\hat{Q}_{sk},\quad
[\hat{H},\hat{Q}_{sk}]\equiv -i\hat{P}_{sk}, \nonumber \\
\hat{P}_{sk}^+=\hat{P}_{sk},\quad
T\hat{P}_{sk}T^{-1}=-\hat{P}_{sk},\quad
[\hat{H},\hat{P}_{sk}]\equiv -i\hat{Q}_{sk}
\end{eqnarray}
and $q_{sk}(t)$ and $p_{sk}(t)$ are the corresponding
T-even and T-odd harmonic deformations given by
\begin{eqnarray}
q_{sk}(t)\equiv <\!\Psi(t)|\hat{Q}_{sk}|\Psi(t)\!>,\quad
p_{sk}(t)\equiv <\!\Psi(t)|\hat{P}_{sk}|\Psi(t)\!>,
\nonumber \\
<\!q_{sk}\!>\equiv <\!\Psi_0|\hat{Q}_{sk}|\Psi_0\!>.
\nonumber \\
\label{6}
\end{eqnarray}
The HF single-particle Hamiltonian is
\begin{eqnarray}
\hat{h}_0(\vec r)  &=&
\sum_{\alpha,s} \frac{\partial {\cal H} (J_{n}^{\alpha},J_{p}^{\alpha})}
{\partial J_{s}^{\alpha}}
\hat{J}_{s}^{\alpha}(\vec r).
\label{7}
\end{eqnarray}
Using the scaling (\ref{4}), it is straightforward to find
time-dependent density and current variations
\begin{eqnarray}
J_{s}^{\alpha}({\vec r},t) && \simeq  J_{s}^{\alpha}({\vec r})+
\delta J_{s}^{\alpha}({\vec r},t), \nonumber \\
\delta J_{s}^{\alpha}({\vec r},t)&&  =
<\Psi(t)|{\hat J}_{s}^{\alpha}|\Psi(t)> -
<\Psi_0|{\hat J}_{s}^{\alpha}|\Psi_0>  =  \nonumber \\
&&  =
-i \sum_k \sum_{s=n,p}
\{(q_{sk}(t)-\!<\!q_{sk}\!>)<|[\hat{P}_{s k},\hat J_{s}^{\alpha}(\vec r)]|>
\nonumber \\
&&  + p_{s k}(t)<|[\hat{Q}_{s k},\hat J_{s}^{\alpha}(\vec r)]|>\}
\label{9}
\end{eqnarray}
and the response Hamiltonian (vibrating single-particle potential)
\begin{eqnarray}
\hat{h}(\vec r,t) && \simeq \hat{h}_0(\vec r) + \hat{h}_{res}(\vec r,t), \nonumber \\
\hat{h}_{res}(\vec r,t) && = \sum_{\alpha's'}
[\frac{\partial {\hat h}_0(\vec r)}{\partial J_{s'}^{\alpha'}}]
\delta J_{s'}^{\alpha'}({\vec r},t)
= \sum_{\alpha s}\sum_{\alpha's'}
[\frac{\partial^2 {\cal H}(\vec r)}
{\delta J_{s}^{\alpha}\delta J_{s'}^{\alpha'}}]
\delta J_{s'}^{\alpha'}({\vec r},t)
{\hat J}_{s}^{\alpha}({\vec r}) \nonumber \\
&& =\sum_{s k}
\{ (q_{sk}(t)-\!<\!q_{st}\!>) \hat{X}_{s k}({\vec r})+
 p_{s k}(t) \hat{Y}_{s k}({\vec r})\}.
\label{10}
\end{eqnarray}
In the above equations we introduced one-body operators
\begin{eqnarray}
\hat{X}_{s k}({\vec r})  && =
i\sum_{s' \alpha' \alpha}
[\frac{\partial^2 {\cal H}} {\partial J_{s'}^{\alpha'}\partial J_{s}^{\alpha}}]
<|[\hat{P}_{s k} ,{\hat J}_{s}^{\alpha}]|>
{\hat J}_{s'}^{\alpha'}(\vec r), \nonumber \\
\hat{Y}_{s k}({\vec r})  && =
i\sum_{s' \alpha' \alpha}
[\frac{\partial^2 {\cal H}} {\partial J_{s'}^{\alpha'}\partial J_{s}^{\alpha}}]
<|[\hat{Q}_{s k} ,{\hat J}_{s}^{\alpha}]|>
{\hat J}_{s'}^{\alpha'}(\vec r)
\label{11}
\end{eqnarray}
where ${\hat X}\!=\!{\hat X}^+$ is T-even, ${\hat Y}\!=\!{\hat Y}^+$ is T-odd,
and the notation
\begin{equation}
<[{\hat A},{\hat B}]>=<\Psi_0|[{\hat A},{\hat B}]|\Psi_0>
\label{12}
\end{equation}
is used. The moments of the operators are
\begin{eqnarray}
<\delta {\hat X}_{s k}(t)>  && \equiv
<\Psi(t)|{\hat X}_{sk}|\Psi(t)>-<\Psi_0|{\hat X}_{sk}|\Psi_0> =
\nonumber \\
&& = i \sum_{s'k'}
(q_{s'k'}(t)-\!<\!q_{s'k'}\!>)<|[\hat{P}_{s'k'},\hat X_{sk}]|> =
\nonumber \\
&& \equiv  \sum_{s'k'} (q_{s'k'}(t)-\!<\!q_{s'k'}\!>) \kappa_{sk,s'k'}^{-1}
\label{13}
\end{eqnarray}
\begin{eqnarray}
<\delta {\hat Y}_{s k}(t)>  && \equiv
<\Psi(t)|{\hat Y}_{sk}|\Psi(t)>-<\Psi_0|{\hat Y}_{sk}|\Psi_0> =
\nonumber \\
&& = i \sum_{s'k'} p_{s'k'}(t) <|[\hat{P}_{s'k'},\hat Y_{sk}]|> =
\nonumber \\
&& \equiv  \sum_{s'k'} p_{s'k'}(t) \eta_{sk,s'k'}^{-1}
\label{14}
\end{eqnarray}
where we introduced the inverse strength matrices
\begin{eqnarray}
\label{15}
\kappa_{sk,s'k'}^{-1 }&&=
\kappa_{s'k',sk}^{-1} =
-i <|[{\hat X}_{sk},\hat{P}_{s'k'}]|> = \\
&& =\int d{\vec r}\sum_{\alpha \alpha'}
[\frac{\delta^2 {\cal H}}{\delta J_{s'}^{\alpha'}\delta J_{s}^{\alpha}}]
<|[{\hat J}_{s}^{\alpha},\hat{P}_{s k}]|>\
<|[{\hat J}_{s'}^{\alpha'},\hat{P}_{s' k'}]|>,\nonumber
\end{eqnarray}
\begin{eqnarray}
\label{16}
\eta_{sk,s'k'}^{-1 }&&=
\eta_{s'k',sk}^{-1 } = -i
<|[{\hat Y}_{sk},\hat{Q}_{s'k'}]|>  \\
&& =\int d{\vec r}\sum_{\alpha \alpha'}
[\frac{\delta^2 {\cal H}} {\delta J_{s'}^{\alpha'}\delta J_{s}^{\alpha}}]
<|[{\hat J}_{s}^{\alpha},\hat{Q}_{s k}]|s >\
<|[{\hat J}_{s'}^{\alpha'},\hat{Q}_{s' k'}]|>.  \nonumber
\end{eqnarray}
Harmonic collective shifts
$q_{sk}(t)-\!\!<\!\!q_{sk}\!\!>$ and velocities $p_{sk}(t)$
read as
\begin{eqnarray}
q_{sk}(t)-\!<\!q_{sk}\!>=\bar{q}_{s k} cos(\omega t)
=\frac{1}{2}\bar{q}_{s k}(e^{i\omega t}+e^{-i\omega t}), \nonumber \\
p_{sk}(t)=\bar{p}_{s k} sin(\omega t)=
\frac{1}{2i}\bar{p}_{s k}(e^{i\omega t}-e^{-i\omega t}).
\label{17}
\end{eqnarray}

Following the Thouless theorem, the perturbed wave function
(\ref{7}) can be also written as
\begin{equation}
|\Psi(t)> \sim (1+\sum_{ph} c_{ph}(t)a^{\dagger}_p a_h) |\Psi_0>
\label{18}
\end{equation}
where $c_{ph}(t)$ are the time-dependent particle-hole contributions
to the given excited state. Substituting (\ref{18}) into the
time-dependent HF equation
\begin{equation}
i\frac{d}{dt}|\Psi(t)>=
({\hat h}_0+{\hat h}_{res}(t))|\Psi(t)>
\label{19}
\end{equation}
and using
\begin{equation}
c_{ph}(t)=c^{+}_{ph}e^{i\omega t}+c^{-}_{ph}e^{-i\omega t},
\label{20}
\end{equation}
one gets the relation between $c^{\pm}_{ph}$ and collective deformations
$\bar{q}_{sk}$ and $\bar{p}_{sk}$
\begin{equation}
c^{\pm}_{ph\epsilon s}=-\frac{1}{2}
\frac{\sum_k [\bar{q}_{sk}
<p|\hat{X}_{sk}|h>
\mp i \bar{p}_{sk}<p|\hat{Y}_{sk}|h>]}
{\varepsilon_{ph}\pm\omega},
\label{21}
\end{equation}
where $\varepsilon_{ph}=\varepsilon_p - \varepsilon_h$,
$\varepsilon_p$ and $\varepsilon_h$ are the particle and hole
energies, respectively.

In addition to Eqs. (\ref{13})-(\ref{14}), the moments
$<\delta {\hat X}_{sk}(t)>$  and $<\delta {\hat Y}_{sk}(t)>$
can be also expressed in terms of the Thouless wave function (\ref{17}):
\begin{eqnarray}
<\!\delta {\hat X}_{sk}(t)\!> && \!\!\!\!\! =  \!\! \sum_{ph}
(c_{ph}(t)^*<\!p|\hat{X}_{sk}|h\!>
+c_{ph}(t)<\!h|\hat{X}_{sk}|p\!>), \nonumber \\
<\!\delta {\hat Y}_{sk}(t)\!> && \!\!\!\!\!  =  \!\! \sum_{ph}
(c_{ph}(t)^*<\!p|\hat{Y}_{sk}|h\!>+
c_{ph}(t)<\!h|\hat{Y}_{sk}|p\!>).
\label{22}
\end{eqnarray}
Then, using Eq. (\ref{21}) an equating the moments  (\ref{13})-(\ref{14})
and (\ref{22}), we obtain the system of equations for unknowns
${\bar q}_{sk}$ and ${\bar p}_{sk}$:
\begin{eqnarray}
\sum_{s'k'} \bar{q}_{s'k'}
[S_{s'k',sk}(XX)-
\kappa_{s'k',sk}^{-1}]+
\sum_{s'k'} \bar{p}_{s'k'}
S_{s'k',sk}(XY)&=&0, \nonumber \\
\sum_{s'k'} \bar{q}_{s'k'}
S_{s'k',sk}(XY)+
\sum_{s'k'} \bar{p}_{s'k'}
[S_{s'k',sk}(YY)-
\eta_{s'k',sk}^{-1}]&=&0
\label{23}
\end{eqnarray}
with
\begin{eqnarray}
S_{s'k',sk}(XX) && =
\sum_{ph}\frac{1}{\varepsilon_{ph}^2-\omega^2}
\{<p|\hat{X}_{s'k'}|h>^* <p|\hat{X}_{sk}|h>
(\varepsilon_{ph}-\omega) \nonumber \\
&& \hspace{1.cm}+
<p|\hat{X}_{s'k'}|h> <p|\hat{X}_{sk}|h>^*
(\varepsilon_{ph}+\omega)\}, \nonumber \\
S_{s'k',sk}(YY) && =
\sum_{ph}\frac{1}{\varepsilon_{ph}^2-\omega^2}
\{<p|\hat{Y}_{s'k'}|>^* <p|\hat{Y}_{sk}|h>
(\varepsilon_{ph}-\omega) \nonumber \\
&& \hspace{1.cm}+
<p|\hat{Y}_{s'k'}|h> <p|\hat{Y}_{sk}|h>^*
(\varepsilon_{ph}+\omega)\}, \nonumber \\
S_{s'k',sk}(XY) && =
-i \sum_{ph}\frac{1}{\varepsilon_{ph}^2-\omega^2}
\{<p|\hat{X}_{s'k'}|>^* <p|\hat{Y}_{sk}|h>
(\varepsilon_{ph}-\omega) \nonumber \\
&& \hspace{1.cm}+
<p|\hat{X}_{s'k'}|h> <p|\hat{Y}_{sk}|h>^*
(\varepsilon_{ph}+\omega)\}.
\label{24}
\end{eqnarray}
The matrix of the system (\ref{23}) is symmetric and real. By
equating the determinant of the system to zero,
$det D=0$, we get the dispersion equation for
eigenvalues $\omega_{\nu}$.

It can be shown that the same equations can be obtained through
the standard RPA equations with the  Hamiltonian
\begin{equation}
{\hat H}_{RPA}={\hat h}_0 + {\hat V}_{res},
\label{26}
\end{equation}
where
\begin{eqnarray}
{\hat V}_{res}= -\frac {1}{2} \sum_{sk} \sum_{s'k'}
[\tilde{\kappa}_{sk,s'k'} {\hat X}_{sk} {\hat X}_{s'k'}
+ \tilde{\eta}_{sk,s'k'} {\hat Y}_{sk} {\hat Y}_{s'k'}].
\label{27}
\end{eqnarray}
The strength constants $\tilde{\kappa}$ and $\tilde{\eta}$  are
obtained by inversion of  the matrices (\ref{15}) and (\ref{16}).

Eqs. (\ref{11}), (\ref{15}), (\ref{16}), (\ref{21}), (\ref{23}), (\ref{24})
constitute the set of SRPA equations.

The rank of SRPA matrix (\ref{23}) is $4K$ with
$K=3-6$ (see next section).  So, high-rank matrix of exact RPA shrinks to low-rank
SRPA matrix. This minimizes the computational effort.
At the same time, as is demonstrated below, SRPA provides high accuracy
of the calculations.

The approach is fully self-consistent and does not need any
adjusting parameters in addition to the starting functional.
Unlike the trivial separable schemes, SRPA gives analytical
expressions for both separable operators and strength constants.
Relative contributions of different separable operators are
self-consistently regulated for every RPA state.

\section{SRPA test with Skyrme functional}

SRPA was tested for the particular case of SkM* functional
\cite{Rei}. Isoscalar and isovector densities (nucleon, kinetic, spin
and spin-orbital) and currents have been involved.
The strength function
 \begin{equation}
  b_L(X\lambda, gr \rightarrow \omega)=\sum_{\nu}\omega_{\nu}^L
 <\nu |\hat{{\cal M}}(X\lambda )|0>^2 \rho(\omega - \omega_{\nu})
\label{35}
 \end{equation}
with Lorentz weight $\rho(x-y)=\frac{1}{2\pi}\frac{\Delta}
{(x-y)^2-(\Delta /2)^2}$ ($\Delta=1$ MeV is the averaging
parameter) has been calculated for isoscalar E2 and isovector E1
transitions in $^{40}Ca$ and $^{208}Pb$ and compared with the
results of exact RPA. In (\ref{35}), $\hat{{\cal M}}(X\lambda )$ is
operator of $X\lambda$-transition.

The set of $K=2$ generators
\begin{equation}
\hat{Q}_{k}({\vec r})= R_k(r)
(Y_{\lambda\mu}+h.c.),
\qquad \hat{P}_k=i[\hat{H},\hat{Q}_k]
\label{41}
\end{equation}
has been used with
$R_1(r)=r^{\lambda}, \quad R_2(r)=r^{\lambda +2}.$
The first $Q_k$-generator coincides with the corresponding
$E\lambda$-transition operator in the long-wave approximation,
and the other creates operators  ${\hat X}_{sk}$ and ${\hat
Y}_{sk}$ with maxima in the interior.

The comparison between SRPA and exact RPA results is exhibited in Fig.
1. Already first operator gives excellent description of E2 giant
resonance. However, it is not enough to reproduce E1 data.
The second operator considerably improves the agreement and
provides a satisfactory description.

\section{Conclusions}

A general procedure for self-consistent factorization of the residual interaction
is proposed for arbitrary  density- and current-dependent
functionals.  The separable RPA (SRPA) constructed in the framework of this
approach dramatically simplifies the calculations while keeping high accuracy
of numerical results. The latter was demonstrated for the case of SkM* functional.

SRPA  can be used for description of $E\lambda$ and $M\lambda$
response in both spherical and deformed nuclei. In the case of
deformed nuclei it allows to take into account the coupling
between electric and magnetic modes. The approach can serve as a
good basis for anharmonic corrections, description of vibrational
states in odd and odd-odd nuclei. SRPA can be especially useful
for investigation of dynamics of deformed nuclei where
applications of effective forces (Skyrme, Gogny, ...) are very
scarce. One of the most promising lines of future studies is
dynamics of exotic nuclei obtained in radioactive beams.

\section*{Acknowledgments}
 This work was partly supported by the Czech grant agency under the contract
No. 202/99/1718 (J.K.). V.O.N.  thanks the support of RFBR (00-02-17194), 
Heisenberg-Landau (Germany-BLTP JINR) and  Votruba-Blokhintcev 
(Czech Republic -BLTP JINR) grants.


\newpage
\noindent
FIGURE CAPTIONS:

\vspace{1cm}
\noindent
{\bf Figure 1}. 
Isoscalar $E2$ and isovector $E1$ strength functions in $^{40}$Ca and 
$^{208}$Pb.
The lines represent exact RPA results (full) and SRPA ones with
one (dotted) and two (dashed) operators.
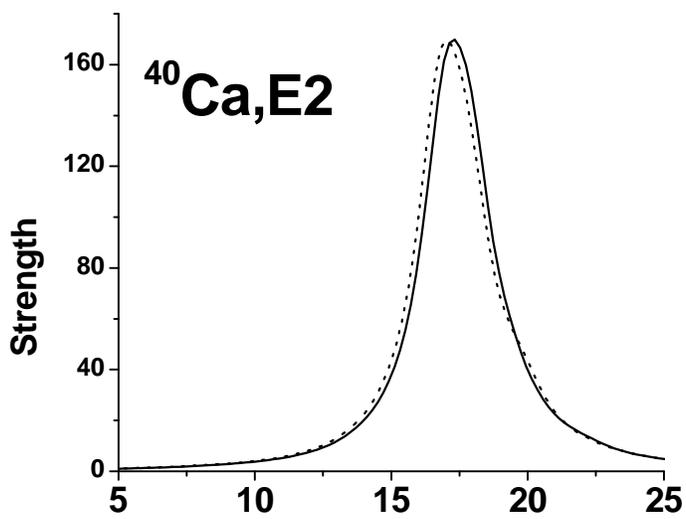
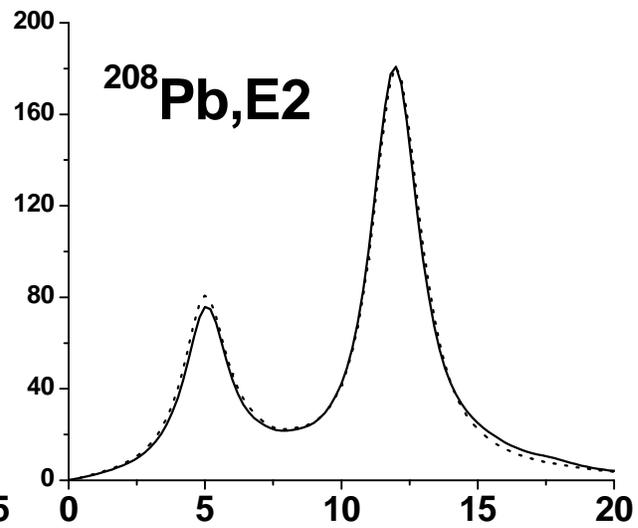
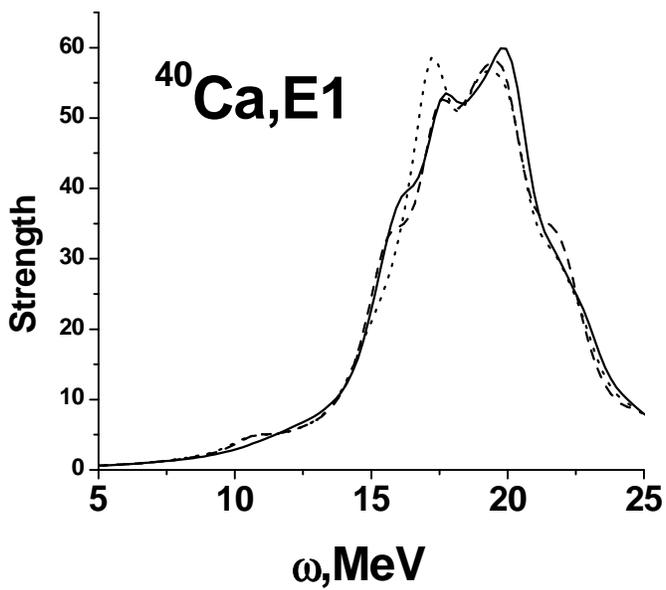
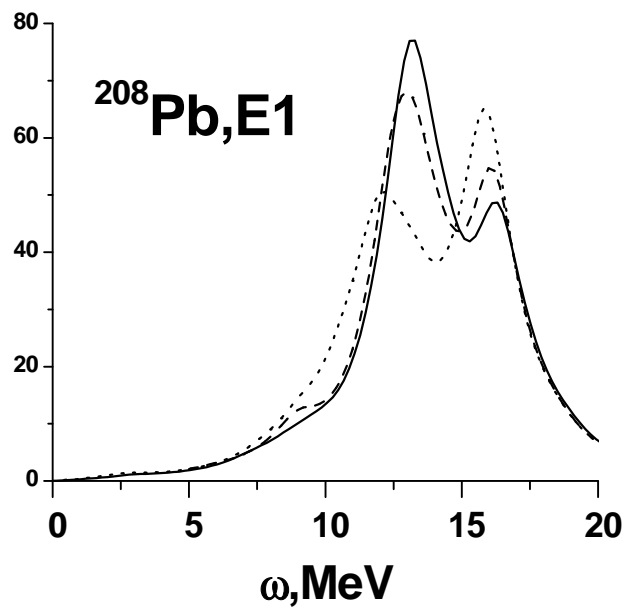
\end{document}